\documentstyle[aps,preprint,eqsecnum,floats,epsf]{revtex}
\tighten
\draft
\begin{document}

\preprint{}
\title{Character Expansions, Itzykson-Zuber Integrals, and the QCD
Partition Function}
\author{A.B. Balantekin\thanks{Electronic address: {\tt
baha@nucth.physics.wisc.edu}}}
\address{Department of Physics, University of Wisconsin\\
         Madison, Wisconsin 53706 USA\thanks{Permanent Address},\\
and\\
Max-Planck-Institut f\"ur Kernphysik,
Postfach 103980, D-69029 Heidelberg, Germany}

\maketitle

\begin{abstract}
A combinatorial formula to generate $U(N)$ character expansions is
presented. It is shown that the resulting character expansion formulas
greatly simplify a number of problems where integrals over the group
manifolds need to be calculated. Several examples are given, including
direct and very quick calculations of the Itzykson-Zuber integral and
the finite volume effective partition function of QCD in the sector
with a given topological charge. 
\end{abstract}

\pacs{}

\newpage

\vglue1cm

\section{Introduction}

Expansion of a periodic function into its Fourier components is widely
used in physics. Since sines and cosines can be considered as the
characters of the $U(1)$ group, Fourier expansion is the simplest
character expansion. In general expansion of an invariant function of
a group into its characters (traces of the representation matrices) is
not easy task. Some time ago the author had given a combinatorial
formula to write character expansions for the $U(N)$ group
\cite{baha1}. The purpose of this paper is first to extend this
formula to more general situations than those covered in
Ref. \cite{baha1}, and then to show that this formula can be
profitably used to simplify a number of situations where integrals
over the group manifolds need to be calculated. 

Derivation of the general character expansion formula is given in the
next section. In Section III we give some examples of character
expansions obtained using our formula. The examples given there are
meant to be illustrative of the technique, but not exhaustive of all
the character expansions one can obtain. Some related determinantal
identities are placed in the Appendix for easy reference. In
Section IV we show that our techniques can be used to directly
calculate the Itzykson-Zuber integral and its various extensions. In
Section V we use our technique to calculate the finite volume
effective partition function of QCD in the sector with a given 
topological charge. Even though the results presented in Sections IV
and V were previously obtained by other methods, it is shown that our
method greatly simplifies the calculations. Finally in Section VI a
brief discussion of the results, including extension of our results
into the $U(N/M)$-type supergroups, and directions for future work 
concludes the paper. For continuity of the text several mathematical
formulae, namely a review of the properties of the symmetric functions
and several determinantal identities are placed in two appendices. 

\section{Character Expansion Formulae}

First we review and extend the main result of
Ref. \cite{baha1}. Consider the representations of the $U(N)$ group
labeled by a partition into $N$ parts: ($n_1,n_2,\cdots,n_N$) where 
$n_1 \ge n_2 \ge \cdots \ge n_N$ (see for example
Ref. \cite{weyl}). The character (trace of the representation matrix) 
of the irreducible representation
corresponding to the partition ($n_1,n_2,\cdots,n_N$) of non-negative
integers is given by Weyl's formula \cite{weyl}:
\begin{equation}
\label{weylsfor1}
\chi_{(n_1,n_2,\cdots,n_N)} (U) = \frac{\det ( t_i^{n_j+N-j})}{\Delta 
(t_1, \cdots, t_N)},
\end{equation}
where $t_i,i=1,\cdots,N$, are the eigenvalues of the group element $U$
in the fundamental representation and $\Delta (t_1, \cdots, t_N)$ is
the Vandermonde determinant in the arguments $t_1, \cdots, t_N$: 
\begin{equation}
\label{vandet}
\Delta (t_1, \cdots, t_N) = \det ( t_i^{N-j}).
\end{equation}
In these equations the arguments of the determinants indicate the
$(ij)$-th element of the matrix the determinant of which is
calculated. An alternative form for the character formula is given by
\begin{equation}
\label{weylsfor2}
\chi_{(n_1,n_2,\cdots,n_N)} (U) = \det ( h_{n_j+i-j}) ,
\end{equation}
where $h_n$ is the complete symmetric function in the arguments $t_1,
\cdots, t_N$ of degree $n$. (For a review of its properties see
Appendix A). 

We now consider the power series expansion
\begin{equation}
\label{powerser}
G(x,t) = \sum_n A_n(x) t^n ,
\end{equation}
where the range of $n$ in the sum is not yet specified. $x$ stands for
all the parameters needed to specify the coefficients $A_n$. We assume
that this series is convergent for $|t|=1$. Given $N$ different $t$'s:
$t_1, \cdots, t_N$, we next write down the equality using $N$ copies
of Eq.(\ref{powerser}) 
\begin{equation}
\label{1}
\Delta (t_1, \cdots, t_N) \left( \prod_{i=1}^N G(x,t_i) \right) = 
\det \left[ \sum_n A_n(x) t_i^{N+n-j} \right]. 
\end{equation}
Changing the variable in the sums to $p=n+N-j$, we can rewrite
Eq.(\ref{1}) as
\begin{equation}
\label{2}
\Delta (t_1, \cdots, t_N) \left( \prod_{i=1}^N G(x,t_i) \right) = 
\det \left[ f_j (t_i) \right],
\end{equation}
where 
\begin{equation}
\label{3}
f_j ( z) = \sum_p A_{p+j-N} z^p .
\end{equation}
The range of $n$ in Eq. (\ref{powerser}), which is so far completely
unrestricted, determines the range of $p$ in Eq. (\ref{3}) . Using the
properties of determinants (see for example Ref. \cite{hua}) and Eq.
(\ref{a2}) of the Appendix B Eq. (\ref{2}) can be written as
\begin{equation}
\label{4}
\Delta (t_1, \cdots, t_N) \left( \prod_{i=1}^N G(x,t_i) \right) =
\sum_{k_1 > k_2 > \cdots > k_N} \det (A_{k_j+i-N}) \det ( t_i^{k_j}). 
\end{equation}
In Eq. (\ref{4}) the range (but not the ordering) of the variables
$k_1, k_2, \cdots, k_N$ are still determined by the range of $n$ in
Eq. (\ref{powerser}). First we introduce 
\begin{equation}
\label{5}
n_i = k_i - N + i. 
\end{equation}
The ordering of the variables indicated in the sum of Eq. (\ref{4})
now becomes  
\begin{equation}
\label{6}
n_1 \ge n_2 \ge \cdots \ge n_N . 
\end{equation}
This transformation is necessary since the partitions $(n_1, \cdots,
n_N)$ that label the representations of $U(N)$ should satisfy the
condition in Eq. (\ref{6}), i.e. it is the $n_i$'s, not the $K_i$'s
that label the irreducible representations. Next we want to show that
because of the condition in Eq. (\ref{6}) only one of the sums in
Eq. (\ref{4}) still spans the original range of $n$ in
Eq. (\ref{powerser}). To this end we introduce the non-negative
quantities 
\begin{equation}
\label{7}
m_j = n_j - n_{j+1}, \> j = 1, \cdots, N-1, 
\end{equation}
one can write
\begin{equation}
\label{8}
n_i = m_i + m_{i+1} + \cdots + m_{N-1} + n_N. 
\end{equation}
As a result the right-hand side of Eq. (\ref{4}) takes the form
\begin{equation}
\label{9}
\sum_{m_1=0} \sum_{m_2 =0} \cdots \sum_{m_{N-1}=0} \sum_{n_N} \det
(A_{n_j +i -j}) \det (t_i^{n_j+N-j}). 
\end{equation}
The upper range of the $m_1$ through $m_{N-1}$ sums are still
determined by the range of $n$ in Eq. (\ref{powerser}), but these sums
start with $m_i=0$. The entire range of the $n_N$ sum is still
determined by the range of $n$ in Eq. (\ref{powerser}).  At this point
we want to use Weyl's formula, Eq. (\ref{weylsfor1}), to write the
last term in the right-hand side of Eq. (\ref{9}) as the
character. Since $n_N$ may take negative values, we cannot yet use
Eq. (\ref{weylsfor1}), where all the $n_j$'s are non-negative. To
achieve our goal we need to do yet another transformation of indices
in Eq. (\ref{9}) to those that take only non-negative values. This is
achieved by introducing the quantities
\begin{mathletters}
\label{10}
\begin{eqnarray}
\ell_i &=& \sum_{j=i}^{N-1} m_j = n_i - n_N, \> i=1,\cdots,N-1  \\
\ell_N &=& 0 . 
\end{eqnarray}
\end{mathletters}
Then the second determinant on the right-hand side of Eq. (\ref{9})
can be written as 
\begin{equation}
\label{11}
\det (t_i^{n_j+N-j}) = \left( t_1 t_2 \cdots t_N \right)^{n_N} \det
(t_i^{\ell_j+N-j}) . 
\end{equation}
Substituting Eq. (\ref{11}) into Eq. (\ref{9}) and then inserting the
resulting expression into Eq. (\ref{4}) one obtains
\begin{eqnarray}
\label{12}
\Delta (&t_1&, \cdots, t_N) \left( \prod_{i=1}^N G(x,t_i) \right)
\nonumber \\ &=& \sum_{m_1=0} \sum_{m_2 =0} \cdots \sum_{m_{N-1}=0}
\sum_{n_N} \det (A_{n_j +i -j}) \left( t_1 t_2 \cdots t_N
\right)^{n_N} \det (t_i^{\ell_j+N-j}) . 
\end{eqnarray}
We can now take $t_i$'s to be the eigenvalues of the fundamental
representation of $U(N)$. Dividing both sides of Eq. (\ref{12}) with
the Vandermonde determinant and using Eq. (\ref{weylsfor1}) we obtain 
\begin{equation}
\label{13}
\left( \prod_{i=1}^N G(x,t_i) \right) = \sum_{m_1=0} \sum_{m_2 =0}
\cdots \sum_{m_{N-1}=0} \sum_{n_N} \det (A_{n_j +i -j}) \left(\det U
\right)^{n_N} \chi_{(\ell_1, \ell_2, \cdots, \ell_N)} (U). 
\end{equation}
This is the main result of this paper. In writing this equation we
used the fact that the Matrix $U$ can always be diagonalized by a
unitary transformation which leaves the character invariant.
Eq. (\ref{13}) is a generalization of the character expansion given in
Ref. \cite{baha1}. If the sum over $n$ in the expression
Eq. (\ref{powerser}) we started with is restricted to the non-negative
values of $n$ (i.e., $A_n=0$ when $n<0$), then $n_N$ is non-negative
and we can absorb the term $(\det U)^{n_N}$ into the character to
obtain the result given in Ref. \cite{baha1}: 
\begin{equation}
\label{14}
\left( \prod_{i=1}^N G(x,t_i) \right) = \sum_{n_1=0} \sum_{n_2 =0}
\cdots \sum_{n_N=0} \det (A_{n_j +i -j}) \chi_{(n_1,n_2,\cdots,n_N)}
(U) .
\end{equation} 
Note that the summation in Eq. (\ref{14}) is over all irreducible
representations of $U(N)$, but in Eq. (\ref{13}) is restricted to
those representations where the number of boxes in the last row of the
Young Tableau is zero {\em and} an additional summation over $n_N$, 
which, in general can take both positive and negative values. An
application of Eq. (\ref{14}) to the thermodynamics of two-dimensional
QCD in the large-N limit was given in Ref. \cite{larry}. In the next
section we give some explicit examples of character expansions. 

\section{Examples of Character Expansions}

For our first example we choose $G(x,t) = \exp (xt)$. Then $A_n=
x^n/n!$ for $n \ge 0$ and $A_n=0$ for $n<0$. We can then use
Eq. (\ref{14}) to write 
\begin{equation}
\label{15}
\exp \left( x Tr U \right) = \sum_r \alpha_r(x) \chi_r (U), 
\end{equation}
where the sum is over all irreducible representations ($r$ stands for
$(n_1,n_2,\cdots,n_N)$) and 
\begin{eqnarray}
\label{16}
\alpha_r (x) &=& \det \left( \frac{x^{n_j+i-j}}{(n_j+i-j)!} \right) =
x^{n_1+n_2+ \cdots + n_N} \nonumber \\
&\times& \left| \begin{array}{cccc} \frac{1}{n_1!} &
\frac{1}{(n_2-1)!} & \frac{1}{(n_3-2)!} & \cdots \\
\frac{1}{(n_1+1)!} & \frac{1}{n_2!} & \frac{1}{(n_3-1)!} & \cdots \\
\frac{1}{(n_1+2)!} & \frac{1}{(n_2+1)!} & \frac{1}{n_3!} & \cdots \\
\vdots & \vdots & \vdots & \ddots
\end{array} \right| . 
\end{eqnarray}
This particular character expansion can also be obtained by explicit
integration over the group manifold \cite{itzhak}. It will be
increasingly difficult to obtain more complicated character expansions
by explicit integration. Using Eqs. (\ref{13}) and (\ref{14})
provides a much easier alternative to the explicit integration over
the group manifold. The determinant in Eq. (\ref{16}) can also be
written in terms of the dimensions of the group representations: 
\begin{equation}
\label{17}
\alpha_{ \{ n_1,n_2,\cdots,n_N \} } = x^{n_1+n_2+ \cdots + n_N}
\left[ \prod_{i=1}^N
\frac{(N-i)!}{(N+n_i-i)!} \right] d_{ \{ n_1,n_2,\cdots,n_N \} }, 
\end{equation}
where $d_{ \{ n_1,n_2,\cdots,n_N \} }$ is the dimension of the
representation corresponding to the partition $\{ n_1,n_2,\cdots,n_N
\}$. (The dimensions can be evaluated by calculating the character of
the identity. See, e.g. Ref. \cite{baha2} for explicit formulas).
A related character expansion can be obtained by noting 
\begin{equation}
\label{17a}
t^{\nu} e^{xt} = \sum_{m=\nu}^{\infty} \frac{x^{m-\nu}}{(m-\nu)!} t^m.
\end{equation}
Using Eq. (\ref{14}) we immediately get 
\begin{equation}
\label{17b}
(\det U)^{\nu} e^{x Tr U} = \sum_r x^{n_1+ \cdots + n_N - N \nu} 
\left( \det \frac{1}{(n_j-\nu +i -j)!} \right) \chi_{n_1,\cdots,n_N}
(U) ,
\end{equation}
and
\begin{equation}
\label{17c}
(\det U)^{- \nu} e^{x Tr U} = \sum_r x^{n_1+ \cdots + n_N + N \nu} 
\left( \det \frac{1}{(n_j+\nu +i -j)!} \right) \chi_{n_1,\cdots,n_N}
(U) .
\end{equation}

For our second example we pick $G(x,t)$ to be the generating function
of the Hermite polynomials:
\begin{equation}
\label{18}
G(x,t) = \exp \left( 2tx - t^2 \right) = \sum_{n=0}^{\infty}
\frac{H_n(x)}{n!} t^n. 
\end{equation}
The corresponding character expansion can again be found using
Eq. (\ref{14}): 
\begin{equation}
\label{19}
\exp \left( a Tr U - b Tr U^2 \right) = \sum_r
b^{(n_1+n_2+\cdots+n_N)/2} \det \left( 
\frac{H_{n_j+i-j}(a/2\sqrt{b})}{(n_j+i-j)!}\right)  \chi_r (U).
\end{equation}

For the next example we choose $G(x,t)$ to be the generating function
of the modified Bessel functions:
\begin{equation}
\label{20}
G(x,t) = \exp \left[ \frac{x}{2} \left( t + \frac{1}{t} \right)
\right] = \sum_{n=-\infty}^{+\infty} I_n(x) t^n .
\end{equation}
Since the index $n$ takes negative as well as positive values we need
to use Eq. (\ref{13}) which yields the character expansion 
\begin{equation}
\label{21}
 \exp \left[ \frac{x}{2} Tr (U + U^{\dagger})
\right] = \sum_{m_1=0}^{+\infty} \sum_{m_2 =0}^{+\infty}
\cdots \sum_{m_{N-1}=0}^{+\infty} \sum_{n_N= - \infty}^{+\infty}
 \det (I_{n_j +i -j}(x)) \left(\det U
\right)^{n_N} \chi_{(\ell_1, \ell_2, \cdots, \ell_N)} (U). 
\end{equation}
This expansion was previously obtained by direct
integration for $SU(N)$ group ($\det U =1$) \cite{itzhak2}.
Because of the symmetry in the argument of the exponential,
Eq. (\ref{21}) can be equivalently written as 
\begin{eqnarray}
\label{21a}
 \exp &&\left[ \frac{x}{2} Tr (U + U^{\dagger}) \right] \nonumber \\
&=& \sum_{m_1=0}^{+\infty} \sum_{m_2 =0}^{+\infty}
\cdots \sum_{m_{N-1}=0}^{+\infty} \sum_{n_N= - \infty}^{+\infty}
 \det (I_{n_j +i -j}(x)) \left(\det U^{\dagger}
\right)^{n_N} \chi_{(\ell_1, \ell_2, \cdots, \ell_N)} (U^{\dagger}).  
\end{eqnarray}
In using these expressions it is useful to remember that $(\det U)^n$
for $n \ge 0$ is the character of the representation where all $n_i =
n, i=1, \cdots,N$: 
\begin{equation}
\label{21b}
(\det U)^n = \chi_{(n,n,\cdots,n)} (U) .
\end{equation}
This can be proven rewriting Eq. (\ref{weylsfor2}) in terms of the
elementary symmetric functions and conjugate partitions; for the
definitions see Appendix A and for a proof see
e.g. Ref. \cite{macdonald}. 

\section{Itzykson-Zuber Integrals}

In 1980 Itzykson and Zuber were able to calculate the group integral
\cite{iz}
\begin{equation}
\label{22}
\int dU \exp \left[ \beta Tr (M_1UM_2U^{\dagger}) \right] =
\left( \prod_{p=o}^{N-1} p! \right) \beta^{-N(N-1)/2} \left[
\frac{\det \left( \exp (\beta \lambda_i \nu_j)
\right)}{\Delta(\lambda)\Delta(\nu)} \right] ,
\end{equation}
where $\lambda$ and $\nu$ are eigenvalues of the matrices $M_1$ and
$M_2$ respectively. This result, which is a special case of a more
general formula by Harish-Chandra \cite{chandra} was extensively used
in the theory of matrix models. Here we present a very simple
direct derivation using the character expansions. 

To derive the Itzykson-Zuber formula using Eq. (\ref{15}) we expand
the integrand 
\begin{equation}
\label{23}
\exp \left[ \beta Tr (M_1UM_2U^{\dagger}) \right] = \sum_r \alpha_r
\chi_r (M_1UM_2U^{\dagger}).
\end{equation}
In writing Eq. (\ref{23}) we assumed that the constant matrices $M_1$
and $M_2$ belong to the group algebra. The group integration is easily
carried out using the formula 
\begin{equation}
\label{24}
\int dU \>\> {\cal U}^{(r)}_{\sigma \beta} \>\> {\cal
U}^{*(r')}_{\gamma \delta} = \frac{1}{d_r} \delta^{r r'}
\delta_{\sigma \gamma} \delta_{\beta \delta}, 
\end{equation}
where  ${\cal U}^{(r)}$ is the group matrix element in the
representation $r$, $d_r$ is the dimension of the representation, and
the Greek indices run from $1$ to $d_r$. A proof of Eq. (\ref{24}) is
given in standard texts, see e.g. Ref. \cite{mur}. Since the character
is  given by $\chi_r (U) = \sum_{\alpha} {\cal U}^{(r)}_{\alpha
\alpha}$ setting $\sigma = \gamma$ and $\beta = \delta$ in
Eq. (\ref{24}) gives the orthogonality formula for the characters:  
\begin{equation}
\label{24a}
\int dU \>\> \chi_r (U) \chi_{r'} (U) =  \delta^{r r'}.
\end{equation} 
Using Eqs.(\ref{23}) and (\ref{24}) one gets
\begin{equation}
\label{25}
\int dU \exp \left[ \beta Tr (M_1UM_2U^{\dagger}) \right] = \sum_r
\frac{\alpha_r}{d_r} \chi_r(M_1) \chi_r(M_2),
\end{equation}
which, using Eq. (\ref{17}) can be written as 
\begin{equation}
\label{26}
\int dU \exp \left[ \beta Tr (M_1UM_2U^{\dagger}) \right] = \sum_r 
\beta^{n_1+n_2+ \cdots + n_N} \left[ \prod_{i=1}^N
\frac{(N-i)!}{(N+n_i-i)!} \right]  \chi_r(M_1) \chi_r(M_2).
\end{equation}
Using Weyl's formula, Eq. (\ref{weylsfor1}), one can rewrite
Eq. (\ref{26}) as
\begin{eqnarray}
\label{27}
\int &dU& \exp \left[ \beta Tr (M_1UM_2U^{\dagger}) \right] \nonumber
\\ &=& \sum_{n_1
\ge n_2 \ge \ldots \ge n_N} \beta^{n_1+n_2+ \cdots + n_N} \left[
\prod_{i=1}^N 
\frac{(N-i)!}{(N+n_i-i)!} \right] \left[
\frac{\det(\lambda_i^{n_j+N-j})
\det(\nu_i^{n_j+N-j})}{\Delta(\lambda)\Delta(\nu)} \right] .
\end{eqnarray}
Replacing $n_i$ by $k_i$ of Eq. (\ref{5}), the above equation takes
the form
\begin{eqnarray}
\label{28}
\int &dU& \exp \left[ \beta Tr (M_1UM_2U^{\dagger}) \right] \nonumber
\\ &=&
\sum_{k_1 > k_2 > \ldots > k_N} \beta^{k_1+k_2+ \cdots + k_N -
N(N-1)/2} \left( \prod_{i=1}^N \frac{(N-i)!}{k_i!} \right) 
\left[ \frac{\det(\lambda_i^{k_j})
\det(\nu_i^{k_j})}{\Delta(\lambda)\Delta(\nu)} \right] .
\end{eqnarray}
Using the power series expansion of the exponential function and the
Theorem Eq. (\ref{a2}) in the Appendix B one can easily rewrite the
right-hand side of Eq. (\ref{28}) to yield the result given in
Eq. (\ref{22}): 
\begin{equation}
\label{28a}
\int dU \exp \left[ \beta Tr (M_1UM_2U^{\dagger}) \right] =
\left( \prod_{p=o}^{N-1} p! \right) \beta^{-N(N-1)/2} \left[
\frac{\det \left( \exp (\beta \lambda_i \nu_j)
\right)}{\Delta(\lambda)\Delta(\nu)} \right] .
\end{equation}

In a later work a generalized form of the Itzykson-Zuber Integral was
calculated \cite{guhr1,jackson}.  The integral in question is  
\begin{equation}
\label{29}
I = \int dU \int dV \exp \left[ i {\Re e} Tr (UxV^{\dagger}y) \right].
\end{equation}
(This integral for complex rectangular ordinary matrices was
previously given in Ref. \cite{berezin}). To calculate this integral
we write 
\begin{equation}
\label{30}
 {\Re e} (UxV^{\dagger}y) = \frac{1}{2} ( U x V^{\dagger} y + y V x
 U^{\dagger} )
\end{equation}
and expand the resulting exponentials using Eq. (\ref{15}):
\begin{eqnarray}
\label{31}
\exp \left[ i {\Re e} Tr (UxV^{\dagger}y) \right] &=& \left[ \sum_r
\left( 
\frac{i}{2} \right)^{n_1 + n_2 + \cdots + n_N} \alpha_r \chi_r ( U x
V^{\dagger} y ) \right] \nonumber \\ &\times& \left[ \sum_{r'} \left(
\frac{i}{2} \right)^{n_1' + n_2' + \cdots + n_N'} \alpha_{r'}
\chi_{r'} ( y V x U^{\dagger} ) \right] 
\end{eqnarray}
Inserting Eq. (\ref{31}) into Eq. (\ref{29}), the group integrals can
easily be carried out using Eq. (\ref{24}) to obtain
\begin{equation}
\label{32}
I = \sum_r \left( \frac{i}{2} \right)^{2(n_1 + n_2 + \cdots + n_N)}
\left[ \frac{\alpha_r^2}{d_r^2} \right] \chi_r (x^2) \chi_r (y^2) . 
\end{equation}
Using Eqs. (\ref{5}), (\ref{17}), and Weyl's character formula this
equation takes the form
\begin{equation}
\label{33}
I = \sum_r \left( \frac{-1}{4} \right)^{n_1 + n_2 + \cdots + n_N}
\frac{\left[ \prod_{i=1}^N (N-i)! \right]^2}{ \left[ \prod_{i=1}^N k_i!
\right]^2}  \left[ \frac{\det(x_i^{2k_j})
\det(y_i^{2k_j})}{\Delta(x^2)\Delta(y^2)} \right] .
\end{equation}
Using the power series expansion of the ordinary Bessel function of
zeroth order 
\begin{equation}
\label{34}
J_o(x) = \sum_{n=0}^{\infty} \frac{ (-1)^n }{ (n!)^2} \left(
\frac{x}{2} \right)^{2n},  
\end{equation}
and the theorem in the Appendix B, Eq. (\ref{33}) can be evaluated to
be 
\begin{equation}
\label{35}
I = (2i)^{N(N+1)} \left[ \sum_{i=0}^{N-1} i! \right]^2 \frac{ \det
\left( J_o (x_iy_j) \right) }{\Delta(x^2)\Delta(y^2)}, 
\end{equation}
which, up to the normalization factors, is the result obtained in
Refs. \cite{guhr1} and \cite{jackson}. 

\section{Effective QCD Partition Functions}

In this section we show that character expansions can be used to
calculate some group integrals which appear in studying effective QCD 
partition functions. The finite volume effective partition function
of QCD in the sector with topological charge $\nu$ can be written as
an integral over $U(N_f)$ \cite{ls}
\begin{equation}
\label{36}
Z_{\nu} = \int dU (\det U)^{\nu} \exp \left[ V \Sigma {\Re e} Tr (
{\cal M} U^{\dagger} ) \right] .
\end{equation}
where $\Sigma$ is a constant related to the value of the quark
condensate in the chiral limit and ${\cal M}$ is the quark mass
matrix. In Ref. \cite{ls} it was shown that for equal quark
masses and $N_f$ flavors this partition function is a particular
determinant with modified Bessel Function entries. A similar form for
different quark masses was conjectured in Ref. \cite{jackson}, but
only proven for $\nu=0$. A proof for different masses in the
case $\nu \neq 0$ was given in Refs. \cite{brower}, \cite{ver1}, and 
\cite{wadati}.  

To calculate the partition function for the case of equal masses we
rewrite it as
\begin{equation}
\label{37}
Z_{\nu} = \int dU (\det U)^{\nu} \exp \left[ \frac{1}{2} V \Sigma m 
Tr ( U + U^{\dagger} ) \right] .
\end{equation}
Using the character expansion in Eq. (\ref{21a}) this takes the form 
\begin{equation}
\label{38}
Z_{\nu} = \int dU \sum_{m_1=0}^{+\infty} \sum_{m_2 =0}^{+\infty}
\cdots \sum_{m_{N-1}=0}^{+\infty} \sum_{n_N= - \infty}^{+\infty}
 \det (I_{n_j +i -j}(V \Sigma m)) \left(\det U
\right)^{\nu - n_N} \chi_{(\ell_1, \ell_2, \cdots, \ell_N)}
(U^{\dagger}), 
\end{equation}
where we used $\det U^{\dagger} = (\det U)^{-1}$. Rewriting $(\det U
)^{\nu - n_N} =  \chi_{(\nu - n_N  ,\nu - n_N, \cdots, \nu - n_N)}
(U)$, and carrying out the group integration using Eq. (\ref{24}) we
get $\nu - n_N = \ell_i = 0$ for all $i$. Eqs. (\ref{7}) and
(\ref{10}) then indicate that the only surviving partition is
$(\nu,\nu, \cdots,\nu)$ and we get 
\begin{equation}
\label{39}
Z_{\nu} =  \det (I_{\nu +i -j}(V \Sigma m)) , 
\end{equation}
which is the desired result. 

For unequal masses we rewrite the partition function as
\begin{equation}
\label{40}
Z_{\nu} = \int dU (\det U)^{\nu} \exp \left[ \frac{1}{2} 
Tr (M^{\dagger} U + U^{\dagger} M ) \right] .
\end{equation}
Note that in this case we cannot use the character expansion in
Eq. (\ref{21a}) since in general $M^{\dagger} U$ is not a unitary
matrix. Instead, using Eq. (\ref{15}) we write
\begin{equation}
\label{41}
\exp \left( x Tr U^{\dagger} M \right) = \sum_r \alpha_r(x) \chi_r
(U^{\dagger} M). 
\end{equation}
Similarly using Eq. (\ref{17b}) we write 
\begin{equation}
\label{42}
(\det M^{\dagger} U)^{\nu} e^{x Tr M^{\dagger} U} = \sum_r x^{n_1+
\cdots + n_N - N \nu} 
\left( \det \frac{1}{(n_j-\nu +i -j)!} \right) \chi_{n_1,\cdots,n_N}
(M^{\dagger} U) .
\end{equation}
Inserting Eqs. (\ref{41}) and (\ref{42}) into Eq. (\ref{40}) the group
integral can easily be done:
\begin{eqnarray}
\label{43}
Z_{\nu} &=& (\det M^{\dagger})^{-\nu} \sum_r \left( \frac{1}{2}
\right)^{2(n_1+ \cdots + n_N) - N \nu}
\left( \det \frac{1}{(n_j - \nu +i -j)!} \right) \nonumber  \\
&\times& \left[
\prod_{i=1}^{N} \frac{ (N-i)!}{(N+n_i -i)!} \right] \chi_r
(M^{\dagger}M) .
\end{eqnarray}
Changing the labels $n_i$ into $k_i$ of Eq. (\ref{5}) and rewriting
$\chi_r (M^{\dagger}M)$ as a ratio of two determinants
(cf. Eq. (\ref{weylsfor1})) we rewrite the partition function as
\begin{eqnarray}
\label{44}
Z_{\nu} &=& (\det M^{\dagger})^{-\nu} \left( \frac{1}{2}
\right)^{-N(N-1) - N \nu} \left[ \prod_{i=1}^{N} (N-i)! \right]
\frac{1}{\Delta (\mu_1^2, \cdots, \mu_N^2)} \nonumber \\
&\times& \sum_{k_1 > k_2 > \cdots > k_N} \det \left[ \frac{1}{k_j!
(k_j - N - \nu +i)!} \right] \det \left[ \left( \frac{\mu_i}{2}
\right)^{2k_j} \right] ,
\end{eqnarray}
where $\mu_i$ are the eigenvalues of the matrix $M$. Noting the power
series expansion of the Bessel Function 
\begin{equation}
\label{45}
\frac{1}{\sqrt{y^{\lambda}}} I_{\lambda} ( 2 \sqrt{y}) =
\sum_{k=0}^{\infty} \frac{1}{k! (\lambda +k)!} y^k ,
\end{equation}
and using Eq. (\ref{a4}) of the Appendix B, Eq. (\ref{44}) takes the
form  
\begin{equation}
\label{46}
Z_{\nu} = \left( \frac{1}{2}
\right)^{-N(N-1)/2} \left[ \prod_{i=1}^{N} (N-i)! \right]
\frac{1}{\Delta (\mu_1^2, \cdots, \mu_N^2)} \det [ \mu_j^{N-i} I_{- \nu
-N +i} (\mu_j) ]
\end{equation}
which is the desired result. Using the fact $I_n = I_{-n}$, the
determinant in the Eq. (\ref{46}) can be rearranged to yield the 
often-quoted form  
\begin{equation}
\label{47}
Z_{\nu} = \left( \frac{1}{2}
\right)^{-N(N-1)/2} \left[ \prod_{i=1}^{N} (N-i)! \right]
\frac{1}{\Delta (\mu_1^2, \cdots, \mu_N^2)} \det [ \mu_i^{j-1} I_{ \nu
+j -1} (\mu_i) ] .
\end{equation}

\section{Conclusions}

Character expansion is a powerful group theoretical technique which
should have more widespread use than it currently enjoys
\cite{migdal}. One obstacle was the difficulty in calculating the
coefficients of the characters in the expansions and the formulas
presented in this paper should help in this regard. We tried to
illustrate the utility of the method by rederiving a number of results
in the literature in a much more direct way. In the future papers we
will cover new applications of the character expansion technique. For
example, in Ref. \cite{gernot} some remarkable relations are derived
among the effective partition functions relevant for describing
microscopic Dirac spectrum. Our method can be used to explore the
nature of such relations and derive new ones. 

The character expansions derived using our formula can be generalized
to the supergroup U(N/M). The characters of the covariant class I
representations of this supergroup are given by a formula similar to
Eq. (\ref{weylsfor2}) except that the complete symmetric functions are
replaced by the graded homogeneous symmetric functions
\cite{baha2}. The former can be written in terms of the traces of the
fundamental representation. The latter are given by similar
expressions except that traces are replaced by supertraces
\cite{baha2,baha3}. Since our character expansion formulas are
basically combinatorial in nature they are applicable in principle to
the covariant representations of the supergroup U(N/M). More general
representations of $U(1/1)$ were considered in Ref. \cite{guhr2}. One
should however note that the characters \cite{baha3} and the invariant
integration \cite{baha4} for the orthosymplectic supergroup
$Osp(N/2M)$ are much more complicated than those of $U(N/M)$. However
an approach based on Gelfand-Tzetlin coordinates may be gainfully
utilized for both $U(N/M)$ and $Osp(N/2M)$ type supergroups
\cite{guhr4}. A detailed analysis of the extension of our character
expansion formula to supergroups will de deferred to later work. 

Another possible application of our formulas is in the random matrix
theory. It was shown that random-matrix theories provide common
concepts to various aspects of quantum phenomena \cite{hans}. Using
random-matrix theory concepts it is possible to study various aspects
of the QCD Dirac operator \cite{guhr3,ver1,ver2}. In this case symmetry
considerations lead to not only chiral unitary, but also chiral
Gaussian orthogonal and symplectic ensembles, which in turn require
generalization of our character expansion formula to orthogonal and
symplectic groups. 

\section*{ACKNOWLEDGMENTS}

I thank T. Guhr, G. Akemann, H. Kohler, and T. Seligman for many
useful 
discussions. This work was supported in part by the U.S. National
Science Foundation Grants No.\ PHY-9605140 and PHY-0070161 
at the University of Wisconsin, in part by the University of Wisconsin
Research Committee with funds granted by the Wisconsin Alumni Research
Foundation, and in part by the Alexander von Humboldt-Stiftung. 
I am grateful to the Max-Planck-Institut f\"ur Kernphysik and
H.A. Weidenm\"uller for the very kind hospitality. 

\section*{Appendix A: Symmetric Functions}

The complete homogeneous symmetric function, $h_n (x)$, of degree $n$
in the arguments $x_i, i=1, \cdots, N$, is defined as the sum of the
products of the variables $x_i$, taking $n$ of them at a time. For
three variables $x_1,x_2,x_3$, the first few complete homogeneous
symmetric functions are
\[
h_1 (x) = x_1 + x_2 + x_3,
\]
\[
h_2(x) = x_1^2 + x_2^2 + x_3^2 + x_1 x_2 +  x_1 x_3 +  x_2 x_3,
\]
\[
h_3 (x) = \sum_i x_i^3 + \sum_{i \neq j} x_i^2 x_j + x_1x_2x_3.
\]
One can write the generating function for $h_n$ as
\begin{equation}
\label{aa1} 
\frac{1}{\prod_{i=1}^N (1 - x_i z)} = \sum_n h_n(x) z^n .
\end{equation}
If $x_1,x_2,x_3$ are the eigenvalues of a matrix $B$, the symmetric
functions can be written in terms of traces of powers of $B$, e. g. 
\[
h_1 (x) = Tr B, 
\]
\[
h_2 (x) = \frac{1}{2} \left[ Tr B^2 + (Tr B)^2 \right],
\]
and so on. 

The elementary symmetric functions, $a_n(x)$, are defined in a similar
way except that no $x_i$ can be repeated in any product. Again for
three variables $x_1,x_2,x_3$, the first few elementary symmetric
functions are 
\[
a_1 = h_1
\]
\[
a_2 = x_1 x_2 +  x_1 x_3 +  x_2 x_3,
\]
\[
a_3 =  x_1x_2x_3.
\]
One takes $a_n = 0$ if $n>N$ and $a_0 = h_0 = 1$. The generating
function for $a_n$ is given by 
\begin{equation}
\label{aa2} 
\prod_{i=1}^N (1 - x_i z) = \sum_n (-1)^n a_n(x) z^n .
\end{equation}
Note that, since the generating functions in Eqs. (\ref{aa1}) and
(\ref{aa2}) are inverses of each other one can write $h_k$ in terms of
$a_i, i=1,\cdots,k$ and vice versa. If one takes $x_i, i=1, \cdots, N$
to be eigenvalues of an $N \times N$ matrix $A$, then $a_N (x) = \det
A$. If $A$ is an element of $U(N)$, then one can consider
Eqs. (\ref{aa1}) and (\ref{aa2}) as character expansions since 
\[
h_n(U) = \chi_{(n,0, \cdots, 0)} (U)
\]
(cf. Eq. (\ref{weylsfor2})) and 
\[
a_N(U) = \chi_{(1,1, \cdots, 1)} (U). 
\]
One can associate a Young Tableau with a given partition $(n_1,
\cdots, n_N)$ where the number of boxes at the $i$th row of the Young
Tableau is $n_i$. A conjugate partition $(m_1,m_2, \cdots, m_N)$ is
defined such that $m_i$ is the number of the boxes at the $i$th
column. One can either write the Weyl's formula using
Eq. (\ref{weylsfor2}) in terms of $h_n$'s as written (in terms of the
partition $(n_1,n_2, \cdots, n_N)$) or in terms of the conjugate
partition by replacing $h_n$'s with $a_n$'s (for details see
Ref. \cite{macdonald}).

\section*{Appendix B: Determinant Expansion Theorems}

Here we state the expansion theorem we use in the text. A proof is
given in Ref. \cite{hua}. Consider the power series expansion
\begin{equation}
\label{a1}
f(z) = a_0 + a_1 z + a_2 z^2 + \cdots ,
\end{equation}
convergent for $|z|< \rho$. Then for $|x_i y_j| < \rho,
i,j=1,\cdots,N$ one can write
\begin{equation}
\label{a2}
\det [ f(x_i y_j) ] = \sum_{k_1 > k_2 > \ldots > k_N} a_{k_1} a_{k_2}
\cdots  a_{k_N}  \det (x_i^{k_j}) \det (y_i^{k_j}) .
\end{equation}
A generalization of this result is known as the Binet-Cauchy
formula. If the power series 
\begin{equation}
\label{a3}
f_i (z) = \sum_{k=0}^{\infty} a^{(i)}_k z^k
\end{equation}
is convergent for $|z|< \rho$, then for $|z_i| < \rho, \forall i$ we
have 
\begin{equation}
\label{a4}
\det f_i (z_j) = \sum_{k_1 > k_2 > \cdots > k_N} (\det a^{(i)}_{k_j})
(\det z_i^{k_j} ) . 
\end{equation}
A proof is given again in  Ref. \cite{hua}.

\end{document}